\documentclass[aps,prl,reprint,superscriptaddress,10pt,a4paper,preprintnumbers,nofootinbib]{revtex4-1}
\usepackage{amsmath,amssymb,latexsym}
\usepackage{bm}
\usepackage{graphicx}
\usepackage{hepunits}
\usepackage{color}

\allowdisplaybreaks

\begin{document}

\title{Mixed QCD-EW corrections for Higgs boson production at $e^+e^-$ colliders}

\author{Yinqiang Gong}
\email{gongyq@pku.edu.cn}
\affiliation{School of Physics and State Key Laboratory of Nuclear Physics and Technology, Peking University, Beijing 100871, China}
\author{Zhao Li}
\email{zhaoli@ihep.ac.cn}
\affiliation{Institute for High Energy Physics, Chinese Academy of Sciences, Beijing 100049, China}
\author{Xiaofeng Xu}
\email{xuxiaofeng@pku.edu.cn}
\affiliation{School of Physics and State Key Laboratory of Nuclear Physics and Technology, Peking University, Beijing 100871, China}
\author{Li Lin Yang}
\email{yanglilin@pku.edu.cn}
\affiliation{School of Physics and State Key Laboratory of Nuclear Physics and Technology, Peking University, Beijing 100871, China}
\affiliation{Collaborative Innovation Center of Quantum Matter, Beijing, China}
\affiliation{Center for High Energy Physics, Peking University, Beijing 100871, China}
\author{Xiaoran Zhao}
\email{xiaoran.zhao@uclouvain.be}
\affiliation{Centre for Cosmology, Particle Physics and Phenomenology (CP3), Université catholique de Louvain, Chemin du Cyclotron, 2, B-1348 Louvain-la-Neuve, Belgium}

\begin{abstract}
Since the discovery of the Higgs boson at the Large Hadron Collider, a future electron-position collider has been proposed for precisely studying its properties. We investigate the production of the Higgs boson at such an $e^+e^-$ collider associated with a $Z$ boson, and calculate for the first time the mixed QCD-electroweak corrections to the total cross sections. We provide an approximate analytic formula for the cross section and show that it reproduces the exact numeric results rather well for collider energies up to $\unit{350}{\GeV}$. We also provide numeric results for $\sqrt{s}=\unit{500}{\GeV}$, where the approximate formula is no longer valid. We find that the $\mathcal{O}(\alpha\alpha_s)$ corrections amount to a $1.3\%$ increase of the cross section for a center-of-mass energy around $\unit{240}{\GeV}$. This is significantly larger than the expected experimental accuracy and has to be included for extracting the properties of the Higgs boson from the measurements of the cross sections in the future.
\end{abstract}

%\pacs{}
%\keywords{}

\maketitle

\section{Introduction}

The discovery of the Higgs boson \cite{Aad:2012tfa, Chatrchyan:2012xdj} at the Large Hadron Collider (LHC) marks a major milestone of high energy physics. Since then, measuring its properties has become a top priority of both experimental and theoretical particle physics. In particular, the knowledge of the couplings of the Higgs field will let us verify the underlying mechanism for electroweak symmetry breaking and the origin of fermion masses, and may shed light on the solution to the hierarchy problem. While the LHC can produce a lot of Higgs bosons, the complicated backgrounds prohibit sufficiently precise measurements to be achieved. For this reason, a next-generation high-luminosity electron-positron collider served as a Higgs factory is undergoing active discussions in the community. Notable candidates include the Circular Electron-Positron Collider (CEPC) \cite{CEPC}, the International Linear Collider (ILC) \cite{Baer:2013cma} and the FCC-ee \cite{Gomez-Ceballos:2013zzn}. Thanks to the clean environment and the high luminosity, many important properties of the Higgs boson can be measured to extremely high precisions at these machines, often orders of magnitude better than the LHC \cite{Peskin:2013xra}. This will provide stringent tests of the Standard Model (SM) and has the potential to indirectly probe new physics beyond the SM which might exist at very high energy scales.

At an electron-positron collider with the center-of-mass energy $\sqrt{s} \sim \unit{240}{\GeV}$, the main production mechanism for the Higgs boson is the associated production of a Higgs boson and a $Z$ boson. At the CEPC, the production cross section $\sigma(e^+e^- \to ZH)$ can be measured to an extremely high precision of $0.51\%$ \cite{CEPC}. As a result, the coupling of the Higgs boson with the $Z$ boson can be extracted with a precision of $0.25\%$. The FCC-ee has claimed an even higher accuracy for this cross section \cite{d'Enterria:2016cpw, d'Enterria:2016yqx}. The studies on the various properties of the Higgs boson rely crucially on detailed investigations of this production channel. Obviously, the theoretical prediction for this process within the SM has to be known with a similar or even higher precision than the experimental one in order for these investigations to be meaningful. The calculations of higher order radiative corrections are therefore indispensable for the Higgs factory projects.

An important feature of the Higgs factories is that the $ZH$ cross section and hence the $HZZ$ coupling can be measured by tagging the leptonically decaying $Z$ boson. In this way one does not need to resort to any particular decay mode of the Higgs boson. Since the branching ratios for the leptonic decays of the $Z$ boson are precisely known, the main theoretical uncertainty comes from our knowledge of the inclusive cross section $\sigma(ZH)$. The next-to-leading order (NLO) electroweak (EW) corrections to $\sigma(ZH)$ have been calculated in \cite{Fleischer:1982af, Kniehl:1991hk, Denner:1992bc}. It was shown that the leading order (LO) cross sections as well as the NLO weak corrections depend crucially on the renormalization scheme used. The total NLO cross sections, on the other hand, are much more stable as expected. For a Higgs mass $m_H \sim \unit{125}{\GeV}$, the NLO cross section reaches its maximum $\sigma(ZH) \sim \unit{230}{\femtobarn}$ for $\sqrt{s} \sim \unit{240}{\GeV}$. This is exactly the design energy of Higgs factories.

Besides the purely weak corrections, there are purely quantum electrodynamics (QED) corrections involving virtual photon exchanges and initial state radiations (ISRs). The QED corrections contain logarithmic enhancements due to soft and collinear photons and can be quite sizable. In \cite{Fleischer:1982af, Kniehl:1991hk, Denner:1992bc}, it was shown that the NLO QED corrections decrease the $ZH$ cross section significantly. The higher order soft photon corrections can be resummed \cite{Gribov:1972rt}, while the hard-collinear corrections have been calculated to the third order \cite{Kuraev:1985hb, Skrzypek:1990qs}. These corrections are often implemented in Monte Carlo event generators. A recent study \cite{Mo:2015mza} using the WHIZARD package \cite{Kilian:2007gr} confirms that the ISR effects reduce $\sigma(ZH)$ by more than 10\% at the CEPC.

Given all these efforts, the theoretical accuracy of the $ZH$ cross section is still not sufficient to match the high experimental precision at Higgs factories, and higher order effects must be considered. In \cite{Kniehl:2012rz}, higher order corrections for a closely related process $H \to ZZ^* \to Zl^+l^-$ were estimated in the large $m_t$ approximation. For $e^+e^- \to ZH$, one needs to go beyond the large $m_t$ limit due to the higher energy scales involved. In this Letter, we take the first step forward by computing exactly the mixed QCD-EW $\mathcal{O}(\alpha\alpha_s)$ corrections to $\sigma(ZH)$, which consist of the main part of the next-to-next-to-leading order (NNLO) corrections for this observable. Our results provide the most precision theoretical predictions for Higgs boson production at Higgs factories, and should be used when extracting the properties of the Higgs boson, in particular the $HZZ$ coupling, from experimental measurements in the future.

\section{Methods}

We consider corrections to the leading order process
\begin{align}
e^-(k_1) + e^+(k_2) \to Z^*(q) \to Z(p_1) + H(p_2) \, ,
\end{align}
and we define $s \equiv Q^2 \equiv q^2$. The leading order cross section can be written as
\begin{align}
\sigma^{\text{LO}} = \frac{2\pi \alpha^2 |\vec{p}_1|/\sqrt{s}}{3\big(s-m_Z^2\big)^2s_w^2c_w^2} \big(E_Z^2+2m_Z^2\big) (v_e^2+a_e^2) \, ,
\end{align}
where $E_Z=(Q^2+m_Z^2-m_H^2)/(2Q)$ is the energy of the on-shell $Z$ boson in the rest frame of $q^\mu$ (i.e., the $e^+e^-$ center-of-mass frame), $s_w = \sin\theta_w$ and $c_w = \cos\theta_w$ with $\theta_w$ the weak mixing angle, $v_e=(-1/4+s_w^2)/(s_wc_w)$ and $a_e=1/(4s_wc_w)$ coming from the vector and axial-vector couplings of the electron with the $Z$ boson, and $|\vec{p}_1|^2=E_Z^2-m_Z^2$.

The $\mathcal{O}(\alpha\alpha_s)$ corrections to the cross section can be decomposed into 3 parts
\begin{align}
\sigma^{\alpha\alpha_s} = \sigma^{\alpha\alpha_s}_Z + \sigma^{\alpha\alpha_s}_\gamma + \sigma^{\alpha\alpha_s}_{eeZ} \, ,
\end{align}
where $\sigma_Z$ consists of corrections to the $HZZ$ vertex (including bubble insertions on the off-shell $Z^*$ propagator), $\sigma_\gamma$ contains contributions from the loop-induced $HZ\gamma^*$ vertex (including the $Z^*$-$\gamma^*$ bubble diagrams), and $\sigma_{eeZ}$ arises from the counter-term for the $eeZ$ vertex. The most general Lorentz structure of the $HZV$ vertices can be written as
\begin{multline}
\mathcal{T}^{\mu\nu}_{HZV} = \frac{iem_Z}{s_wc_w} \, \bigg( p_1^\mu p_1^\nu T_{1,V} + q^\mu q^\nu T_{2,V} - p_1^\mu q^\nu T_{3,V}
\\
- \frac{q^\mu p_1^\nu}{QE_Z} T_{4,V} + g^{\mu\nu} T_{5,V} - \epsilon^{\mu\nu\rho\sigma} p_{1\rho} q_{\sigma} T_{6,V} \bigg) \, .
\end{multline}
Among the coefficients $T_{i,V}$ in the above formula, only $T_{4,V}$ and $T_{5,V}$ contribute to the unpolarized cross section:
\begin{multline}
\sigma^{\alpha\alpha_s}_V = \frac{4\pi \alpha^2 |\vec{p}_1|/\sqrt{s}}{3\big(s-m_Z^2\big)s_w^2c_w^2} \, C_V
\\
\times \big[ \big(E_Z^2+2m_Z^2\big) \, \mathrm{Re} T^{\alpha\alpha_s}_{5,V} - \big(E_Z^2-m_Z^2\big) \, \mathrm{Re} T^{\alpha\alpha_s}_{4,V} \big] \, ,
\end{multline}
where $C_Z=(v_e^2+a_e^2)/(s-m_Z^2)$ and $C_\gamma=v_e/s$.
The contribution from the counter-term for the $eeZ$ vertex is given by
\begin{multline}
\label{eq:eeZ}
\sigma^{\alpha\alpha_s}_{eeZ} = \frac{4\pi \alpha^2 |\vec{p}_1|/\sqrt{s}}{3\big(s-m_Z^2\big)^2s_w^2c_w^2} \big( E_Z^2+2m_Z^2 \big) \Bigg[ \frac{\delta Z^{\alpha\alpha_s}_{\gamma Z}}{2} v_e + 
\\
\bigg( \frac{\delta Z^{\alpha\alpha_s}_{ZZ}}{2} + \delta Z^{\alpha\alpha_s}_{e} \bigg) \big(v_e^2+a_e^2\big) + v_e \delta v^{\alpha\alpha_s}_e + a_e \delta a^{\alpha\alpha_s}_e \Bigg] \, ,
\end{multline}
where $\delta v_e = -\delta c_w^2 (1+2s_w^2)/(8s_w^3c_w^3)$ and $\delta a_e = \delta c_w^2 (1-2s_w^2)/(8s_w^3c_w^3)$.

For the calculation of $T_{4,V}$ and $T_{5,V}$, we generate the relevant Feynman diagrams using both FeynArts \cite{Hahn:2000kx} and QGRAF \cite{Nogueira:1991ex}. The resulting amplitudes are further manipulated with FORM \cite{Vermaseren:2000nd}. The 2-loop integrals arising from the triangle diagrams with a top quark loop are reduced to a minimal set of 41 master integrals using Reduze 2 \cite{vonManteuffel:2012np} and FIRE \cite{Smirnov:2014hma}, which implement the Laporta algorithm \cite{Laporta:2001dd} for solving integrate by parts (IBP) identities. Part of these master integrals, together with those in the bubble diagrams and those in the calculation of renormalization constants, can be calculated analytically using the method of differential equations \cite{Henn:2014qga}. The remaining master integrals do not admit analytic solutions and we evaluate them with two different methods.

The first method involves an series expansion in $1/m_t$. This expansion breaks down once the center-of-mass energy goes above the $t\bar{t}$ threshold. We therefore expect that this method is valid for $\sqrt{s} < 2m_t$. The leading contribution to the $\mathcal{O}(\alpha\alpha_s)$ correction is of order $m_t^2$, and we perform the expansion up to order $m_t^{-4}$. In this way we obtain an approximate analytic formula for the cross section. This approximate result is to be compared with the result from a purely numerical evaluation of the difficult master integrals using the method of sector decomposition \cite{Binoth:2000ps, Binoth:2003ak}. For this second method we have used a fast private code documented in \cite{Li:2015foa}, and cross-checked with SecDec \cite{Borowka:2015mxa} whenever possible.

We renormalize the fields and the masses in the on-shell scheme. The weak mixing angle is defined by the on-shell relation $c_w=m_W/m_Z$, whose renormalization constant is given by $\delta c_w^2/c_w^2 = \delta m_W^2/m_W^2 - \delta m_Z^2/m_Z^2$. For the renormalization of the fine structure constant $\alpha$, we choose to work in schemes which are insensitive to non-perturbative effects and to the masses of light fermions such that we can safely take them to zero. Specifically, we show results in two different schemes. The first scheme involves renormalizing $\alpha$ in the $\overline{\text{MS}}$ scheme for all contributions except the top quark loop, which is subtracted on-shell. In this scheme the fine structure constant becomes scale-dependent and we denote it by $\hat{\alpha}(\mu)$. An alternative scheme is to subtract the low-energy contributions due to light fermions from the on-shell renormalized $\alpha(0)$, and define an effective coupling $\alpha(m_Z) = \alpha(0)/\big(1-\Delta\alpha(m_Z)\big)$. For a review of these schemes and the recent evaluations of the hadronic contributions to $\Delta\alpha(m_Z)$, see \cite{Agashe:2014kda}.

As mentioned before, the benefit of performing the expansion in $1/m_t$ is that we obtain an approximate analytical formula for the cross section, which allows much faster numerical evaluations compared to the sector decomposition method. For our numerical results in the next section we have used the expansion up to order $m_t^{-4}$. Due to the limited space, we give below the analytic results up to order $m_t^0$, which will prove to be a sufficiently accurate approximation for $\sqrt{s} \sim \unit{250}{\GeV}$. We begin with the simpler ones:
\begin{align}
T_{4,\gamma}^{\alpha\alpha_s} &= \frac{\alpha}{4\pi} \frac{\alpha_s}{4\pi} C_F \frac{8QE_Z v_t}{m_Z^2} + \mathcal{O}(m_t^{-2}) \, ,
\\
T_{5,\gamma}^{\alpha\alpha_s} &= \frac{\alpha}{4\pi} \frac{\alpha_s}{4\pi} C_F \Bigg[ \frac{8QE_Zv_t}{m_Z^2} 
\\
&- \frac{\big(21-44s_w^2\big) Q^2}{3s_wc_w(Q^2-m_Z^2)} \bigg( \ln\frac{Q^2}{m_Z^2} + i\pi \bigg) \Bigg] + \mathcal{O}(m_t^{-2}) \, , \nonumber
\\
T_{4,Z}^{\alpha\alpha_s} &= \frac{\alpha}{4\pi} \frac{\alpha_s}{4\pi} C_F \frac{QE_Z}{m_Z^2} \left( -12 v_t^2 + \frac{4}{3} a_t^2 \right) + \mathcal{O}(m_t^{-2}) \, ,
\end{align}
where $v_t=(1/4-2s_w^2/3)/(s_wc_w)$ and $a_t=-1/(4s_wc_w)$ come from the vector and axial-vector couplings of the top quark with the $Z$ boson. Note that all the above 3 coefficients vanish at the leading order. The most complicated coefficient is $T_{5,Z}$, which equals 1 at tree-level. It is given by
\begin{align}
\label{eq:T5Z}
T_{5,Z}^{\alpha\alpha_s} &= \frac{\alpha}{4\pi} \frac{\alpha_s}{4\pi} C_F \Bigg\{ \frac{m_t^2}{m_Z^2} a_t^2 \big( 30 - 12\pi^2 - 264L_t - 144L_t^2 \big) \nonumber
\\
&\hspace{-2em} + \frac{\big(45-84s_w^2+88s_w^4\big)}{6s_w^2c_w^2(Q^2-m_Z^2)} \Bigg[ m_Z^2 + Q^2 \bigg( \ln\frac{Q^2}{m_Z^2} + i\pi - 1 \bigg) \Bigg] \nonumber
\\
&\hspace{-2em} - 12(v_t^2+a_t^2) \frac{Q E_Z}{m_Z^2} - \frac{4}{3} a_t^2 \frac{m_H^2}{m_Z^2}  \Bigg\} + \mathcal{O}(m_t^{-2})
\\
&\hspace{-2em} + \Bigg[ \delta Z_e + \delta Z_{ZZ} + \frac{1}{2} \delta Z_H + \frac{\delta m_Z^2}{2m_Z^2} + \frac{\delta c_w^2 (c_w^2-s_w^2)}{2s_w^2c_w^2} \Bigg]_{\text{finite}}^{\alpha\alpha_s} , \nonumber
\end{align}
where $L_t = \ln(\mu^2/m_t^2)$, and the subscript ``finite'' refers to the finite part of the various renormalization constants.

The renormalization constants appearing in Eqs.~(\ref{eq:eeZ}) and (\ref{eq:T5Z}) are calculated exactly with the help of differential equations. We have checked that our results agree with those in \cite{Djouadi:1993ss, Djouadi:1994gf}.\footnote{There is a typo in \cite{Djouadi:1993ss} relevant for the $W^\pm$ boson self-energies, which was corrected in \cite{Dittmaier:2015rxo}.}
%Since we have used a different renormalization scheme for the electromagnetic coupling, we list its renormalization constant here:
%\begin{align}
%\delta Z_e^{\alpha\alpha_s} = \frac{\alpha}{4\pi} \frac{\alpha_s}{4\pi} C_F \Bigg( \frac{5}{\epsilon} + 10 + \frac{8}{3} L_t \Bigg) \, .
%\end{align}

\section{Results}

In this section we present the numerical predictions from our calculations. We choose the input parameters as $m_t=\unit{173.3}{\GeV}$, $m_H=\unit{125.1}{\GeV}$, $m_Z=\unit{91.1876}{\GeV}$, $m_W=\unit{80.385}{\GeV}$, $\hat{\alpha}(m_Z)=1/127.94$, $\alpha(m_Z) = 1/128.933$ and $\alpha_s(m_Z)=0.118$ \cite{Agashe:2014kda}. The default renormalization scale is chosen as $\mu_0 = \sqrt{s}/2$. The renormalization group evolutions of the coupling constants are performed at 4 loops for $\hat{\alpha}$ \cite{Erler:1998sy} and 2 loops for $\alpha_s$. We calculate the NLO weak corrections using FeynArts \cite{Hahn:2000kx} and FormCalc \cite{Hahn:1998yk}.

\begin{table}[t!]
\centering
\begin{tabular}{c|c|c|c|c}
\hline
$\sqrt{s}$ (GeV) & $\sigma_{\text{LO}}$ (fb) & $\sigma_{\text{NLO}}$ (fb) & $\sigma_{\text{NNLO}}$ (fb) & $\sigma_{\text{NNLO}}^{\text{exp.}}$ (fb)
\\ \hline
240 & 256.3(9) & 228.0(1) & 230.9(4) & 230.9(4)
\\ \hline
250 & 256.3(9) & 227.3(1) & 230.2(4) & 230.2(4)
\\ \hline
300 & 193.4(7) & 170.2(1) & 172.4(3) & 172.4(3)
\\ \hline
350 & 138.2(5) & 122.1(1) & 123.9(2) & 123.6(2)
\\ \hline
500 & 61.38(22) & 53.86(2) & 54.24(7) & 54.64(10)
\\ \hline
\end{tabular}
\caption{\label{tab:xs}Total cross sections at various collider energies in the $\overline{\text{MS}}$ scheme.}
\end{table}

\begin{table}[t!]
\centering
\begin{tabular}{c|c|c|c|c}
\hline
$\sqrt{s}$ (GeV) & $\sigma_{\text{LO}}$ (fb) & $\sigma_{\text{NLO}}$ (fb) & $\sigma_{\text{NNLO}}$ (fb) & $\sigma_{\text{NNLO}}^{\text{exp.}}$ (fb)
\\ \hline
240 & 252.0 & 228.6 & 231.5 & 231.5
\\ \hline
250 & 252.0 & 227.9 & 230.8 & 230.8
\\ \hline
300 & 190.0 & 170.7 & 172.9 & 172.9
\\ \hline
350 & 135.6 & 122.5 & 124.2 & 124.0
\\ \hline
500 & 60.12 & 54.03 & 54.42 & 54.81
\\ \hline
\end{tabular}
\caption{\label{tab:xsAMZ}Total cross sections at various collider energies in the $\alpha(m_Z)$ scheme.}
\end{table}

% \begin{table}[h!]
% \centering
% \begin{tabular}{c|c|c|c|c}
% \hline
% $\sqrt{s}$ (GeV) & $\sigma_{\text{LO}}$ (fb) & $\sigma_{\text{NLO}}$ (fb) & $\Delta\sigma_{\alpha\alpha_s}$ (fb) & $\Delta\sigma_{\alpha\alpha_s}^{\text{expasion}}$ (fb)
% \\ \hline
% 240 & 255.93 & 228.02 & 3.10 & 3.09
% \\ \hline
% 250 & 255.92 & 227.33 & 3.10 & 3.10
% \\ \hline
% 300 & 192.94 & 170.25 & 2.41 & 2.37
% \\ \hline
% 350 & 137.69 & 122.14 & 1.97 & 1.73
% \\ \hline
% 500 & 61.057 & 53.887 & 0.466 & 0.919
% \\ \hline
% \end{tabular}
% \caption{Reference numbers for $\mu = m_Z$.}
% \end{table}

In Table~\ref{tab:xs} we show the NNLO predictions along with the LO and the NLO cross sections in the $\overline{\text{MS}}$ scheme for center-of-mass energies $\sqrt{s}=\unit{240}{\GeV}$, $\unit{250}{\GeV}$, $\unit{300}{\GeV}$, $\unit{350}{\GeV}$ and $\unit{500}{\GeV}$. The results from the $1/m_t$ expansion up to order $m_t^{-4}$ are also shown in the 5th column in the table. In Table~\ref{tab:xsAMZ}, we show the same information, but in the $\alpha(m_Z)$ scheme. We find that the $\mathcal{O}(\alpha\alpha_s)$ corrections increase the NLO cross section by about $1.3\%$ for all 3 collider energies below the $t\bar{t}$ threshold of about $\unit{346}{\GeV}$. This effect is significantly larger than the expected experimental accuracies of Higgs factories. Our results are therefore crucial for extracting theoretical parameters from precision measurements at these future facilities. We also see that the $1/m_t$ expansion approximates the exact results remarkably well for these 3 energies. The digits in the parentheses reflect the variations of the cross sections with respect to the renormalization scale $\mu$ by a factor of 2 around the default scale $\mu_0=\sqrt{s}/2$. We observe that the variations of the NLO cross sections are too small to cover the higher order corrections, which is common for electroweak observables. The mixed QCD-EW corrections introduce dependence on strong interactions for the first time in the perturbative series. As a result, the NNLO cross sections exhibit larger scale variations than the NLO ones. Comparing Table~\ref{tab:xs} and \ref{tab:xsAMZ}, one can see that the results in the two schemes are quite close to each other. For the NNLO results, the difference between the two schemes are similar in size to the effect of scale variation in the $\overline{\text{MS}}$ scheme. We use these to give a rough estimate that the size of even higher order corrections amounts to about 0.2\%.

Once we go for higher energies above the $t\bar{t}$ threshold, the $1/m_t$ expansion is expected to break down. In this case one has to rely on the numerical methods. Nevertheless, we observe from Table~\ref{tab:xs} and \ref{tab:xsAMZ} that for $\sqrt{s}=\unit{350}{\GeV}$, the $1/m_t$ expansion still does a reasonable job to describe the $\mathcal{O}(\alpha\alpha_s)$ correction. We also see that, due to the threshold enhancement, the NNLO correction can reach 1.5\% of the NLO cross section. The energy $\sqrt{s}=\unit{350}{\GeV}$ is just slightly above the $t\bar{t}$ threshold\footnote{This fact also makes the numerical evaluation of the master integrals for $\sqrt{s}=\unit{350}{\GeV}$ rather difficult. For this reason, many optimizations over the original version of the program reported in \cite{Li:2015foa} are implemented to further improve the efficiency. We are not able to cross-check this result using the current public version of SecDec (3.0.9) with the computation resource attainable to us.}, and is a design energy of the ILC and the FCC-ee to study the properties of the top quark, which makes it particularly interesting. Our result provides the essential theoretical input to continue investigating the Higgs boson at this collider energy. 

Going further up to higher energies, the main task of the colliders becomes producing new particles below the TeV scale rather than precisely measuring standard model processes, and the $ZH$ cross section is not as important as in previous cases. Nevertheless, we give the results for $\sqrt{s}=\unit{500}{\GeV}$ in Table~\ref{tab:xs} and \ref{tab:xsAMZ} for demonstration purposes. It is clear that the asymptotic expansion completely fails here: the $1/m_t$ expansion up to order $m_t^{-4}$ overestimates the size of the NNLO correction by a factor of 2.

To further assess the behavior of the $1/m_t$ expansion, we show in Table~\ref{tab:exp} the fractions of different orders of the expansion in the full $\mathcal{O}(\alpha\alpha_s)$ corrections at the default scale $\mu=\sqrt{s}/2$ in the $\overline{\text{MS}}$ scheme. Results in the $\alpha(m_Z)$ scheme are similar and we do not show them here. Again we show the results for 5 different center-of-mass energies. The most important one is $\sqrt{s}=\unit{240}{\GeV}$, which exhibits the largest production cross section and also very high luminosity can be achieved experimentally, and therefore is the design energy of Higgs factories. At this energy, we see that the leading $\mathcal{O}(m_t^2)$ term accounts for about $82\%$ of the total corrections, while the subleading $\mathcal{O}(m_t^0)$ term accounts for another $16\%$. The even higher power contributions are negligible here. These demonstrate the good convergence of the $1/m_t$ expansion and the usefulness of our approximate analytical formula, which evaluates much faster than the sector decomposition method. It provides an efficient and reliable way to perform high precision physics analyses for Higgs factories.

As we increase the center-of-mass energy, it can be seen that the size of the power corrections starts to grow gradually. The $1/m_t$ expansion still provides very good approximations to the full results as long as the energies are below or even slightly above the $t\bar{t}$ threshold. For $\sqrt{s}=\unit{500}{\GeV}$ which is far beyond the threshold, the power series tends to diverge as expected.

\begin{table}[t!]
\centering
\begin{tabular}{c|c|c|c|c}
\hline
$\sqrt{s}$ (GeV) & $\mathcal{O}(m_t^2)$ & $\mathcal{O}(m_t^0)$ & $\mathcal{O}(m_t^{-2})$ & $\mathcal{O}(m_t^{-4})$
\\ \hline
240 & 81.8\% & 16.2\% & 1.4\% & 0.4\%
\\ \hline
250 & 81.7\% & 16.1\% & 1.5\% & 0.5\%
\\ \hline
300 & 80.0\% & 15.2\% & 2.1\% & 1.1\%
\\ \hline
350 & 69.7\% & 12.6\% & 2.7\% & 2.1\%
\\ \hline
500 & 137\% & 18.6\% & 17.3\% & 31.1\%
\\ \hline
\end{tabular}
\caption{\label{tab:exp}Convergence of the $1/m_t^2$ expansion for the mixed QCD-EW corrections in the $\overline{\text{MS}}$ scheme with $\mu=\sqrt{s}/2$.}
\end{table}

\section{Summary and outlook}

In this Letter, we calculated the mixed QCD-electroweak corrections to the associated production of a Higgs boson and a $Z$ boson at future electron-positron colliders. We found that the $\mathcal{O}(\alpha\alpha_s)$ corrections increase the cross sections by about $1.3\%$, which is significantly larger than the expected experimental accuracies of the Higgs factories. Our results should be used when extracting the properties of the Higgs boson, in particular the $HZZ$ coupling, from future precision measurements of the $ZH$ production cross section. While we only presented our predictions for the total cross sections in this Letter, it is rather straightforward to use our formula to study the kinematic distributions as well as polarized scatterings with high precisions.

We have shown that for center-of-mass energies below the $t\bar{t}$ threshold, the approximate analytic formula obtained in the $1/m_t$ expansion agrees remarkably well with the exact numeric results. This is especially important for the design energy of the Higgs factories $\sqrt{s} \sim \unit{240}{\GeV}$, as it provides a fast and reliable method to perform physics analyses with high precisions. We have also shown that even for $\sqrt{s} = \unit{350}{\GeV}$, which is slightly above the $t\bar{t}$ threshold and is a design energy of the ILC and the FCC-ee to study the top quarks, the approximate formula is still valid with good precisions. For higher collider energies which can be achieved at the ILC and the FCC-ee, such an expansion breaks down for $\sqrt{s}$ much larger than $2m_t$, and we have explicitly demonstrated that for $\sqrt{s}=\unit{500}{\GeV}$.

The mixed QCD-EW corrections calculated in this Letter brings the accuracy of the theoretical prediction for the $ZH$ cross section to a regime comparable to the expected experimental accuracy at Higgs factories. At this point, one should further consider several other important effects. First of all, as we stated in the Introduction, one should add the QED corrections upon our results. Secondly, in addition to the production process, one should combine the NNLO corrections to the decay of the on-shell $Z$ boson to leptons and also consider the corrections from the finite width of the $Z$ boson. Finally, it is interesting to know the size of the two-loop weak corrections of order $\alpha^2$. Such a calculation is unlikely to be done with analytic methods. However, with the recent developments in the numerical evaluation of loop integrals, a purely numerical estimation of the remaining NNLO effects should be possible in the near future. 

\vspace{1ex}

{\em Acknowledgments:\/}
This work was supported in part by the National Natural Science Foundation of China under Grant No. 11575004, No. 11635001, No. 11305179, No. 11675185, and by the European Union as part of the FP7 Marie Curie Initial Training Network MCnetITN (PITN-GA-2012-315877).

\vspace{1ex}

Note added after submission: while we are finishing this work, we are aware of an independent calculation \cite{Sun:2016bel} for the same process.


\begin{thebibliography}{99}

%\cite{Aad:2012tfa}
\bibitem{Aad:2012tfa} 
  G.~Aad {\it et al.} [ATLAS Collaboration],
  %``Observation of a new particle in the search for the Standard Model Higgs boson with the ATLAS detector at the LHC,''
  Phys.\ Lett.\ B {\bf 716}, 1 (2012)
  [arXiv:1207.7214 [hep-ex]].

%\cite{Chatrchyan:2012xdj}
\bibitem{Chatrchyan:2012xdj} 
  S.~Chatrchyan {\it et al.} [CMS Collaboration],
  %``Observation of a new boson at a mass of 125 GeV with the CMS experiment at the LHC,''
  Phys.\ Lett.\ B {\bf 716}, 30 (2012)
  [arXiv:1207.7235 [hep-ex]].

\bibitem{CEPC} 
  CEPC-SPPC Study Group,
  CEPC-SPPC Preliminary Conceptual Design Report.

%\cite{Baer:2013cma}
\bibitem{Baer:2013cma} 
  H.~Baer {\it et al.},
  %``The International Linear Collider Technical Design Report - Volume 2: Physics,''
  arXiv:1306.6352 [hep-ph].

%\cite{Gomez-Ceballos:2013zzn}
\bibitem{Gomez-Ceballos:2013zzn} 
  M.~Bicer {\it et al.} [TLEP Design Study Working Group Collaboration],
  %``First Look at the Physics Case of TLEP,''
  JHEP {\bf 1401}, 164 (2014)
  [arXiv:1308.6176 [hep-ex]].

%\cite{Peskin:2013xra}
\bibitem{Peskin:2013xra} 
  M.~E.~Peskin,
  %``Estimation of LHC and ILC Capabilities for Precision Higgs Boson Coupling Measurements,''
  arXiv:1312.4974 [hep-ph].
  
%\cite{d'Enterria:2016cpw}
\bibitem{d'Enterria:2016cpw} 
  D.~d'Enterria,
  %``Physics case of FCC-ee,''
  Frascati Phys.\ Ser.\  {\bf 61}, 17 (2016)
  [arXiv:1601.06640 [hep-ex]].
  
%\cite{d'Enterria:2016yqx}
\bibitem{d'Enterria:2016yqx} 
  D.~d'Enterria,
  %``Physics at the FCC-ee,''
  arXiv:1602.05043 [hep-ex].

%\cite{Fleischer:1982af}
\bibitem{Fleischer:1982af} 
  J.~Fleischer and F.~Jegerlehner,
  %``Radiative Corrections to Higgs Production by $e^+ e^- \to Z H$ in the {Weinberg-Salam} Model,''
  Nucl.\ Phys.\ B {\bf 216}, 469 (1983).
  
%\cite{Kniehl:1991hk}
\bibitem{Kniehl:1991hk} 
  B.~A.~Kniehl,
  %``Radiative corrections for associated $Z H$ production at future $e^{+} e^{-}$ colliders,''
  Z.\ Phys.\ C {\bf 55}, 605 (1992).
  
%\cite{Denner:1992bc}
\bibitem{Denner:1992bc} 
  A.~Denner, J.~Kublbeck, R.~Mertig and M.~Bohm,
  %``Electroweak radiative corrections to e+ e- ---> H Z,''
  Z.\ Phys.\ C {\bf 56}, 261 (1992).

%\cite{Gribov:1972rt}
\bibitem{Gribov:1972rt} 
  V.~N.~Gribov and L.~N.~Lipatov,
  %``e+ e- pair annihilation and deep inelastic e p scattering in perturbation theory,''
  Sov.\ J.\ Nucl.\ Phys.\  {\bf 15}, 675 (1972)
  [Yad.\ Fiz.\  {\bf 15}, 1218 (1972)].

%\cite{Kuraev:1985hb}
\bibitem{Kuraev:1985hb} 
  E.~A.~Kuraev and V.~S.~Fadin,
  %``On Radiative Corrections to e+ e- Single Photon Annihilation at High-Energy,''
  Sov.\ J.\ Nucl.\ Phys.\  {\bf 41}, 466 (1985)
  [Yad.\ Fiz.\  {\bf 41}, 733 (1985)].
  
%\cite{Skrzypek:1990qs}
\bibitem{Skrzypek:1990qs} 
  M.~Skrzypek and S.~Jadach,
  %``Exact and approximate solutions for the electron nonsinglet structure function in QED,''
  Z.\ Phys.\ C {\bf 49}, 577 (1991).

%\cite{Mo:2015mza}
\bibitem{Mo:2015mza} 
  X.~Mo, G.~Li, M.~Q.~Ruan and X.~C.~Lou,
  %``Physics cross sections and event generation of $e^+e^-$ annihilations at the CEPC,''
  Chin.\ Phys.\ C {\bf 40}, no. 3, 033001 (2016)
  [arXiv:1505.01008 [hep-ex]].

%\cite{Kilian:2007gr}
\bibitem{Kilian:2007gr} 
  W.~Kilian, T.~Ohl and J.~Reuter,
  %``WHIZARD: Simulating Multi-Particle Processes at LHC and ILC,''
  Eur.\ Phys.\ J.\ C {\bf 71}, 1742 (2011)
  [arXiv:0708.4233 [hep-ph]].
  
%\cite{Kniehl:2012rz}
\bibitem{Kniehl:2012rz} 
  B.~A.~Kniehl and O.~L.~Veretin,
  %``Low-mass Higgs decays to four leptons at one loop and beyond,''
  Phys.\ Rev.\ D {\bf 86}, 053007 (2012)
  [arXiv:1206.7110 [hep-ph]].
  
%\cite{Hahn:2000kx}
\bibitem{Hahn:2000kx} 
  T.~Hahn,
  %``Generating Feynman diagrams and amplitudes with FeynArts 3,''
  Comput.\ Phys.\ Commun.\  {\bf 140}, 418 (2001)
  [hep-ph/0012260].
  
%\cite{Nogueira:1991ex}
\bibitem{Nogueira:1991ex} 
 P.~Nogueira,
 %``Automatic Feynman graph generation,''
 J.\ Comput.\ Phys.\  {\bf 105}, 279 (1993).

%\cite{Vermaseren:2000nd}
\bibitem{Vermaseren:2000nd} 
 J.~A.~M.~Vermaseren,
 %``New features of FORM,''
 math-ph/0010025.
 
%\cite{vonManteuffel:2012np}
\bibitem{vonManteuffel:2012np} 
 A.~von Manteuffel and C.~Studerus,
 %``Reduze 2 - Distributed Feynman Integral Reduction,''
 arXiv:1201.4330 [hep-ph].

%\cite{Smirnov:2014hma}
\bibitem{Smirnov:2014hma} 
 A.~V.~Smirnov,
 %``FIRE5: a C++ implementation of Feynman Integral REduction,''
 Comput.\ Phys.\ Commun.\  {\bf 189}, 182 (2015)
 [arXiv:1408.2372 [hep-ph]].

%\cite{Laporta:2001dd}
\bibitem{Laporta:2001dd} 
  S.~Laporta,
  %``High precision calculation of multiloop Feynman integrals by difference equations,''
  Int.\ J.\ Mod.\ Phys.\ A {\bf 15}, 5087 (2000)
  [hep-ph/0102033].

%\cite{Henn:2014qga}
\bibitem{Henn:2014qga} 
  J.~M.~Henn,
  %``Lectures on differential equations for Feynman integrals,''
  J.\ Phys.\ A {\bf 48}, 153001 (2015)
  [arXiv:1412.2296 [hep-ph]].

%\cite{Binoth:2000ps}
\bibitem{Binoth:2000ps} 
  T.~Binoth and G.~Heinrich,
  %``An automatized algorithm to compute infrared divergent multiloop integrals,''
  Nucl.\ Phys.\ B {\bf 585}, 741 (2000)
  [hep-ph/0004013].
  
%\cite{Binoth:2003ak}
\bibitem{Binoth:2003ak} 
  T.~Binoth and G.~Heinrich,
  %``Numerical evaluation of multiloop integrals by sector decomposition,''
  Nucl.\ Phys.\ B {\bf 680}, 375 (2004)
  [hep-ph/0305234].
  
%\cite{Li:2015foa}
\bibitem{Li:2015foa} 
  Z.~Li, J.~Wang, Q.~S.~Yan and X.~Zhao,
  %``Efficient numerical evaluation of Feynman integrals,''
  Chin.\ Phys.\ C {\bf 40}, no. 3, 033103 (2016)
  [arXiv:1508.02512 [hep-ph]].
  
%\cite{Borowka:2015mxa}
\bibitem{Borowka:2015mxa} 
  S.~Borowka, G.~Heinrich, S.~P.~Jones, M.~Kerner, J.~Schlenk and T.~Zirke,
  %``SecDec-3.0: numerical evaluation of multi-scale integrals beyond one loop,''
  Comput.\ Phys.\ Commun.\  {\bf 196}, 470 (2015)
  [arXiv:1502.06595 [hep-ph]].

 
%\cite{Agashe:2014kda}
\bibitem{Agashe:2014kda} 
  K.~A.~Olive {\it et al.} [Particle Data Group Collaboration],
  %``Review of Particle Physics,''
  Chin.\ Phys.\ C {\bf 38}, 090001 (2014).
  
%\cite{Djouadi:1993ss}
\bibitem{Djouadi:1993ss} 
  A.~Djouadi and P.~Gambino,
  %``Electroweak gauge bosons selfenergies: Complete QCD corrections,''
  Phys.\ Rev.\ D {\bf 49}, 3499 (1994)
  Erratum: [Phys.\ Rev.\ D {\bf 53}, 4111 (1996)]
  [hep-ph/9309298].

%\cite{Djouadi:1994gf}
\bibitem{Djouadi:1994gf} 
  A.~Djouadi and P.~Gambino,
  %``QCD corrections to Higgs boson selfenergies and fermionic decay widths,''
  Phys.\ Rev.\ D {\bf 51}, 218 (1995)
  Erratum: [Phys.\ Rev.\ D {\bf 53}, 4111 (1996)]
  [hep-ph/9406431].

%\cite{Erler:1998sy}
\bibitem{Erler:1998sy} 
  J.~Erler,
  %``Calculation of the QED coupling alpha (M(Z)) in the modified minimal subtraction scheme,''
  Phys.\ Rev.\ D {\bf 59}, 054008 (1999)
  [hep-ph/9803453].
  
%\cite{Hahn:2000kx}
\bibitem{Hahn:2000kx} 
  T.~Hahn,
  %``Generating Feynman diagrams and amplitudes with FeynArts 3,''
  Comput.\ Phys.\ Commun.\  {\bf 140}, 418 (2001)
  [hep-ph/0012260].

%\cite{Hahn:1998yk}
\bibitem{Hahn:1998yk} 
  T.~Hahn and M.~Perez-Victoria,
  %``Automatized one loop calculations in four-dimensions and D-dimensions,''
  Comput.\ Phys.\ Commun.\  {\bf 118}, 153 (1999)
  [hep-ph/9807565].

%\cite{Dittmaier:2015rxo}
\bibitem{Dittmaier:2015rxo} 
  S.~Dittmaier, A.~Huss and C.~Schwinn,
  %``Dominant mixed QCD-electroweak O($\alpha$$_s$$\alpha$) corrections to Drell–Yan processes in the resonance region,''
  Nucl.\ Phys.\ B {\bf 904}, 216 (2016)
  [arXiv:1511.08016 [hep-ph]].


%\cite{Sun:2016bel}
\bibitem{Sun:2016bel} 
  Q.~F.~Sun, F.~Feng, Y.~Jia and W.~L.~Sang,
  %``Mixed electroweak-QCD corrections to $e^+e^-\to HZ$ at Higgs factories,''
  arXiv:1609.03995 [hep-ph].
  
\end{thebibliography}
\end{document}